\documentclass[aip,pop,reprint]{revtex4-1}

\usepackage{amsmath}
\usepackage{graphicx}% Include figure files
\usepackage{bm}% bold math
\usepackage{url}

\begin{document}

\title{Band Structure of the Growth Rate of the Two-Stream Instability of an
Electron Beam Propagating in a Bounded Plasma }
\author{I. D. Kaganovich$^{a}$}
\author{D. Sydorenko$^{b}$}
\affiliation{$^{a}$Princeton Plasma Physics Laboratory, Princeton University,
Princeton, New Jersey 08543, USA \\
$^{b}$University of Alberta, Edmonton, Alberta T6G 2E1, Canada }

\begin{abstract}
This paper presents a study of the two-stream instability of an electron beam
propagating in a finite-size plasma placed between two electrodes. It is
shown that the growth rate in such a system is much smaller than that of an
infinite plasma or a finite size plasma with periodic boundary conditions.
Even if the width of the plasma matches the resonance condition for a
standing wave, a spatially growing wave is excited instead with the growth
rate small compared to that of the standing wave in a periodic system. The
approximate expression for this growth rate is $\gamma \approx (1/13)\omega
_{pe}(n_{b}/n_{p})(L\omega _{pe}/v_{b})\ln (L\omega _{pe}/v_{b})\left[
1-0.18\cos \left( L\omega _{pe}/v_{b}+{\pi }/{2}\right) \right]$, where
$\omega _{pe}$ is the electron plasma frequency, $n_{b}$ and $n_{p}$ are the
beam and the plasma densities, respectively, $v_{b}$ is the beam velocity,
and $L$ is the plasma width. The frequency, wave number and the spatial and
temporal growth rates as functions of the plasma size exhibit band structure.
The amplitude of saturation of the instability depends on the system length,
not on the beam current. For short systems, the amplitude may exceed values
predicted for infinite plasmas by more than an order of magnitude.
\end{abstract}

\maketitle

% Beam-plasma instabilities, 52.35.Qz
% Beams, interactions with plasma, 52.40.Mj
% Plasma simulation, 52.65.-y
% Plasma applications, 52.77.-j

% also possible:
% Accelerated beams, in plasmas, 52.59.Bi, 52.59.Fn
% Particle beams, intense beams in plasma, 52.59.-f
% Plasma waves, electrostatic waves and oscillations, 52.35.Fp

\section{Introduction \label{sec:intro}}

Interaction of electron beams with plasmas is of considerable importance for
many plasma applications where electron emission occurs from surfaces. The
electrons extracted from the surface and accelerated in the sheath form a beam
of electrons; the beam propagating in the plasma excites electron plasma waves
through the two-stream instability.\cite{Briggs:1964} Laboratory plasmas and
plasmas in industrial applications are usually bounded by electrodes. We show
that electrodes greatly affect the growth of the two-stream instability. Though
beam-plasma systems have been studied extensively in the past using kinetic
simulations,\cite{Kainer:1972,Morey:1989,Gunell:1996,Yoon:2005} the presence of
electrically connected boundaries changes the character of the two-stream
instability from convective to absolute, similar to the instability of a Pierce
diode.\cite{Pierce:1944} In the Pierce diode, the instability was studied
extensively taking only beam electrons and neutralizing ions into account as
relevant to vacuum diodes, see \textit{e.g.} Ref.~\onlinecite{Piel:2010} and
the references within. Here, we consider the two-stream instability between a
low density electron beam and high density plasma electrons as relevant to
discharges. In this Letter, we have performed an analytical study and fluid and
particle-in-cell simulations in order to obtain the growth rate of the
two-stream instability in a finite plasma bounded by electrically connected
electrodes. To the best of our knowledge and to some extent to our surprise the
solution to this problem was not reported before.

The linear stage of the instability can be described making use of fluid
formalism which includes the continuity equations
%======================================================================= eq:01
\begin{equation}\label{eq:01}
\frac{\partial n_{e,b}}{\partial t}+\frac{\partial v_{e,b}n_{e,b}}{\partial x%
}=0,
\end{equation}%
the momentum equations
%======================================================================= eq:02
\begin{equation} \label{eq:02}
\frac{\partial v_{e,b}}{\partial t}+v_{e,b}\frac{\partial v_{e,b}}{\partial x%
}=-\frac{e}{m}E,
\end{equation}%
and the Poisson equation
%======================================================================= eq:03
\begin{equation}\label{eq:03}
\frac{\partial ^{2}\phi }{\partial x^{2}}=4\pi e\left(
n_{e}+n_{b}-n_{i}\right),
\end{equation}%
where $n_{e,b}$ and $v_{e,b}$ are the densities and the velocities of the
plasma and beam electrons, $-e$ and $m$ are the electron charge and mass, $%
E=-{\partial \phi }/{\partial x}$ is the electric field, $\phi$ is the
electric potential, and $n_{i}$ is the ion density. The initial plasma state
is neutral: $n_{e,0}+n_{b,0}=n_{i,0}$, where $n_{e,0}$ and $n_{b,0}$ are the
initial densities of the bulk and the beam electrons, and $n_{i,0}$ is the
initial density of ions, respectively. The ion density is uniform and
constant, $n_{i}=n_{i,0}=const$. Initially, the bulk and the beam electron
densities and the beam flow velocity are uniform everywhere. Note that
everywhere in this paper subscripts $e$ and $b$ denote values related to plasma
and beam electrons, respectively.

For the studies described in the present paper, the boundary conditions are
non-periodic and describe a plasma produced in a discharge between two
electrodes. At the ends of the system $x=0$ and $x=L$, the potential
perturbations are set to zero, $\phi (0)=\phi (L)=0$. The beam is injected at
the boundary $x=0$. The boundary conditions for the beam electrons are
$n_{b}(0)=n_{b,0}$ and $v_{b}(0)=v_{b,0}$, where $v_{b,0}$ is the injection
velocity of the beam. Note that in fluid simulations, a small sheath forms near
the electrodes and more accurate boundary conditions are required to account
for the sheath effect.\cite{Smolyakov:2013}

The paper is organized as follows. In Section~\ref{sec:analytical}, a
dispersion relation for the finite-length beam-plasma system is derived. In
Section~\ref{sec:fluid}, complex frequencies and wavenumbers obtained by direct
solution of the dispersion equation are compared with the fluid simulation and
approximate analytical formulas for the frequency, wavenumber, and temporal and
spatial growth rates are given. Section~\ref{sec:kinetic} compares growth rates
in kinetic simulations with the predictions of the fluid theory.
Section~\ref{sec:kinampl} discusses the amplitude of saturation of the
instability and provides analytical formulas for the estimate of the saturation
electric field amplitude. The results are summarized in
Section~\ref{sec:summary}.

\section{Analytical Solution \label{sec:analytical}}

The dispersion equation is obtained by solving linearized
Eqs.~(\ref{eq:01}-\ref{eq:03}) for perturbations of plasma and beam electron
densities and velocities. The perturbations are defined as
%===============================================================================
\begin{equation*}
\begin{split}
\delta {n_{e}}& =n_{e}-n_{e,0},~\delta {n_{b}}=n_{b}-n_{b,0}, \\
\delta {v_{b}}& =v_{b}-v_{b,0},~\delta {v_{e}}=v_{e}.
\end{split}
\end{equation*}
Linearized equations can be readily solved using Laplace's
method.\cite{Rosenbluth:1963} However, we are only looking for an asymptotic
solution which the system approaches on longer times. Following the Pierce
method,\cite{Pierce:1944} the asymptotic solution for the potential has the
following form:
%========================================================================= eq:04
\begin{equation}\label{eq:04}
\delta \phi (t,x)=\left( {Ax+Be^{ik_{+}x}+Ce^{ik_{-}x}+D}\right) e^{-i\omega
t},
\end{equation}%
where $\omega $ is the frequency of the wave, $k_{\pm }$ are the wave
vectors of the two waves propagating in the system, and coefficients $A,B,C,D
$ are complex constants. The density and the velocity perturbations are
%========================================================================= eq:05
\begin{equation}\label{eq:05}
\begin{split}
\delta {n_{e,b}}(t,x)& =\left( \delta {n_{e,b}^{\prime }}+\delta {n_{e,b}^{+}%
}e^{ik_{+}x}+\delta {n_{e,b}^{-}}e^{ik_{-}x}\right) e^{-i\omega t}, \\
\delta {v_{e,b}}(t,x)& =\left( \delta {v_{e,b}^{\prime }}+\delta {v_{e,b}^{+}%
}e^{ik_{+}x}+\delta {v_{e,b}^{-}}e^{ik_{-}x}\right) e^{-i\omega t},
\end{split}
\end{equation}%
The linearized equations for the parts of the perturbations proportional to $%
\exp (-{i}\omega {t}+ik_{\pm }x)$ are
%===============================================================================
\begin{equation*}
\begin{split}
-i\omega \delta n_{e}^{\pm }+ik_{\pm }\delta v_{e}^{\pm }n_{e,0}& =0, \\
-i\omega \delta v_{e}^{\pm }& =\frac{e}{m}ik_{\pm }\delta \phi ^{\pm }, \\
-i\omega \delta n_{b}^{\pm }+ik_{\pm }\left( {\delta v_{b}^{\pm
}n_{b,0}+v_{b,0}\delta n_{b}^{\pm }}\right) & =0, \\
\left( {-i\omega +ik_{\pm }v_{b,0}}\right) \delta v_{b}^{\pm }& =\frac{e}{m}%
ik_{\pm }\delta \phi ^{\pm }, \\
-k_{\pm }^{2}\delta \phi ^{\pm }& =4\pi e\left( {\delta n_{e}^{\pm }+\delta
n_{b}^{\pm }}\right) ,
\end{split}
\end{equation*}
where $\delta\phi^+=B$ and $\delta\phi^-=C$. These equations yield:
%========================================================================= eq:06
\begin{equation}\label{eq:06}
\begin{split}
\delta v_{e}^{\pm }& =\frac{\omega }{k}\frac{\delta n_{e}^{\pm }}{n_{e,0}},~%
\frac{\delta n_{e}^{\pm }}{n_{e,0}}=-\frac{e}{m}\frac{k_{\pm }^{2}}{\omega
^{2}}\delta \phi ^{\pm }, \\
\delta v_{b}^{\pm }& =\frac{\omega -v_{b,0}k_{\pm }}{k_{\pm }}\frac{\delta
n_{b}^{\pm }}{n_{b,0}},~\frac{\delta n_{b}^{\pm }}{n_{b,0}}=-\frac{e}{m}%
\frac{k_{\pm }^{2}}{\left( {\omega -kv_{b}^{\pm }}\right) ^{2}}\delta \phi
^{\pm }.
\end{split}
\end{equation}%
Substitution of relations (\ref{eq:06}) into the Poisson equation gives
usual dispersion relation for waves
%========================================================================= eq:07
\begin{equation}\label{eq:07}
1=\frac{\omega _{e,0}^{2}}{\omega ^{2}}+\frac{\omega _{b,0}^{2}}{(\omega
-v_{b,0}k_{\pm })^{2}}.
\end{equation}
Here $\omega _{e,0}^{2}\equiv 4\pi e^{2}n_{e,0}/m$ and $\omega
_{b,0}^{2}\equiv 4\pi e^{2}n_{b,0}/m$ are the electron plasma frequencies
corresponding to the plasma and beam densities.

The uniform parts of the density and velocity perturbations (\ref{eq:05}),
which are proportional to $\exp (-i\omega t)$ and correspond to
high-frequency uniform electric field given by the first term in Eq.~(\ref%
{eq:04}), are obtained in a similar way:
%========================================================================= eq:08
\begin{equation}\label{eq:08}
\delta v_{e}^{\prime }=\delta v_{b}^{\prime }=\frac{ie}{\omega m}A,\delta
n_{e}^{\prime }=\delta n_{b}^{\prime }=0~.
\end{equation}%
These perturbations correspond to high-frequency current flowing through the
plasma and allow for $\delta v_{e}\neq 0$ at the systems ends;
$\delta v_{b}(0)=0$ because beam is injected with a given velocity but $\delta
{v}_{b}(L)\neq 0$.

Applying four boundary conditions $\delta{n}_b(0)=\delta{v}_{b}(0)
=\delta\phi(0)=\delta\phi(L)=0$ to perturbations (\ref{eq:04}) and (\ref{eq:05})
and taking into account (\ref{eq:06}) and (\ref{eq:07}) in the form
%========================================================================= eq:09
\begin{equation}\label{eq:09}
\omega-k_\pm v_{b,0}=\pm\frac{\omega_{b,0}} {\sqrt{1-\dfrac{\omega_{e,0}^2}
{\omega^2}}}
\end{equation}
gives the following additional relation between $\omega $ and $k$:
%========================================================================= eq:10
\begin{equation}\label{eq:10}
\begin{split}
k_-^2\left({e^{ik_+ L}-1}\right)- \frac{ik_-^2k_+ \omega L} {\omega -k_+
v_{b,0}} = \\
k_+^2\left({e^{ik_- L}-1}\right)- \frac{ik_+^2k_-\omega L}{\omega-k_- v_{b,0}}.
\end{split}%
\end{equation}
Eqs.~(\ref{eq:09}) and (\ref{eq:10}) determine the temporal
$[\text{Im}(\omega)]$
and the spatial $[\text{Im}(k)]$ growth rates of the instability as well as
the frequency $[\text{Re}(\omega)]$ and the wavenumber $[\text{Re}({k})]$. If
plasma electrons are absent and only beam electrons are taken into account
$(n_{e,0}=0)$, Eq.~(\ref{eq:10}) reduces to the Pierce's dispersion relation
for vacuum diode.

In order to solve the dispersion relation (\ref{eq:10}), we introduce a new
dimensionless variable
%========================================================================= eq:11
\begin{equation}\label{eq:11}
\chi=\frac{\omega_{b,0}/\omega_{e,0}} {\sqrt{1-\dfrac{\omega_{e,0}^2} {%
\omega^2}}}.
\end{equation}
Substituting (\ref{eq:11}) into (\ref{eq:09}) and assuming that $%
\omega=\omega_{e,0}$ in the left-hand side of (\ref{eq:09}) gives
%========================================================================= eq:12
\begin{equation}\label{eq:12}
k_\pm =(1\mp \chi)\frac{\omega_{e,0}}{v_{b,0}}.
\end{equation}
Substitution (\ref{eq:12}) into (\ref{eq:10}) yields equation for $\chi$
%========================================================================= eq:13
\begin{equation}\label{eq:13}
\begin{split}
-i\frac{2(1-\chi)}{(1+\chi)\chi}L_{n} +e^{i(1-\chi)L_{n}}-1- \\
\frac{(1-\chi)^{2}}{(1+\chi)^{2}}\left[{e^{i(1+\chi)L_{n}}-1} \right]=0~,
\end{split}
\end{equation}
where $L_n \equiv L\omega_{e,0} /v_{b,0}$ is the normalized gap width.

Equation (\ref{eq:13}) gives $\chi$ as a function of $L_{n}$. The frequency is
calculated from (\ref{eq:11}) and for a low-density beam with $n_{b,0}\ll
n_{e,0}$ it is
%========================================================================= eq:14
\begin{equation}\label{eq:14}
\omega =\frac{\omega_{e,0}} {\sqrt{1-\dfrac{\omega_{b,0}^2}{\omega_{e,0}^2
\chi^2}}} \approx \omega_{e,0}\left({1+\frac{n_{b,0}}{2n_{e,0}\chi^2}}%
\right).
\end{equation}
The wavenumbers $k_\pm$ can be obtained from (\ref{eq:12}).

Function $\chi (L_{n})$ is complex with band structure, \textit{i.e.} it
changes abruptly at certain $L_{n}=c+2\pi l$, where $c$ is a constant and $l$
is an integer. Indeed, in the limit of $L_{n}\gg 1$, the first two terms in
(\ref{eq:13}) are dominant which gives the following approximate expression:
%===============================================================================
\begin{equation*}
-i{\chi }L_{n}e^{-i{\chi }L_{n}}=2L_{n}^{2}e^{-iL_{n}}.
\end{equation*}
Here, we used the fact that $|\chi|\ll 1$ and $\text{Im}(\chi)>0$. The solution
of this equation is the Lambert or productlog function:~\cite{Corless:1996}
%========================================================================= eq:15
\begin{equation}\label{eq:15}
-i{\chi }L_{n}=W\left( {2L_{n}^{2}e^{-iL_{n}}}\right) .
\end{equation}
This function has many branches, the branch selected must ensure the maximal
growth rate. When parameters of the plasma, \textit{e.g.} the discharge gap,
change, a transition from one branch to another may occur and the
instability growth rate will change abruptly.

Since $\chi$ is complex and independent on $n_{b,0}$, it follows from
(\ref{eq:14}) that the temporal growth rate
of the instability is proportional to $\omega_{e,0}(n_{b,0}/n_{e,0})$ unlike
the growth rate of the resonant perturbation $k\approx
\omega_{e,0} /v_{b,0} $ in a periodic system proportional to
$\omega_{e,0} \left( {n_{b,0} /n_{e,0} } \right)^{1/3}$ \cite{Briggs:1964}.

The analytical solution is verified by fluid and particle-in-cell (PIC)
simulations described below.

\section{Fluid simulations \label{sec:fluid}}

The fluid numerical model solves Eqs.~(\ref{eq:01})-(\ref{eq:03}). The
densities in (\ref{eq:01}) are advanced using the SHASTA method.\cite{SHASTA:1}
The velocities in (\ref{eq:02}) are advanced using an upwind
scheme.\cite{upwind:1}
The model demonstrates excellent agreement with the theory \cite%
{Briggs:1964} in simulations of the instability of a cold beam in a cold
plasma with periodic boundary conditions.

The fluid simulations are carried out with the following common parameters:
$n_{e,0}=2\times{10}^{17}\text{~m}^{-3}$,
$\omega_{e,0}=2.52\times{10}^{11}\text{~s}^{-1}$, beam energy 50 eV and beam
velocity $v_{b,0}=4.2\times{10}^6\text{~m/s}$, the numerical grid cell size is
$1.3\text{~}\mu\text{m}$, the time step is $0.9\text{~fs}$. The selected values
of spatial and temporal steps ensure stability of the SHASTA algorithm. The
resonant beam wavelength $\lambda_b\equiv 2\pi v_{b,0}/\omega_{e,0} $ is 1.044
mm for these plasma parameters. Initially, the bulk electron flow velocity is
given a harmonic perturbation $\delta v_{e}=\delta
v_{e,0}\sin(x\omega_{e,0}/v_{b,0})$ with the wavelength corresponding to the
resonance in a periodic or an infinite plasma, the amplitude of the perturbation
is very small, $\delta v_{e,0}=0.1\text{~m/s}$.

\begin{figure}
\includegraphics{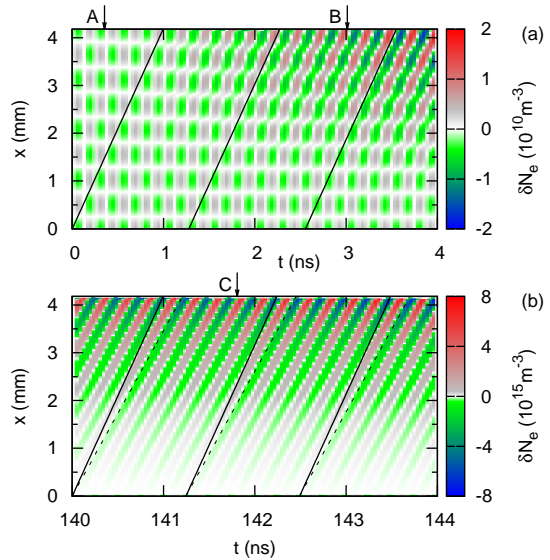}
\caption{\label{fig:01} %
Evolution of the bulk electron density perturbation in time and space in fluid
simulation with $L=4\lambda_b$ and $\alpha=0.0006$. Panels (a) and (b)
correspond to the very beginning of the fluid simulation (a) and to the
asymptotic constant growth stage (b); the corresponding temporal growth of the
electric field amplitude is shown by the red curve in Fig.~\protect\ref{fig:03}.
Solid black lines in (a) and (b) represent propagation with the unperturbed beam
velocity. Dashed black lines in (b) represent phase velocity of the wave
calculated as $\text{Re}(\omega)/\text{Re}(k)$, where
$\text{Re}(\omega)=2.522\times{10}^{10}\text{~s}^{-1}$ and
$\text{Re}(k)=7.288\text{~mm}^{-1}$. Arrows A, B, and C mark times
$t_A=0.35\text{~ns}$, $t_B=3.01\text{~ns}$, and $t_C=141.8\text{~ns}$ when
profiles shown in Figs.~\ref{fig:02}a, \ref{fig:02}b, and \ref{fig:02}c are
obtained. }
\end{figure}

\begin{figure}
\includegraphics{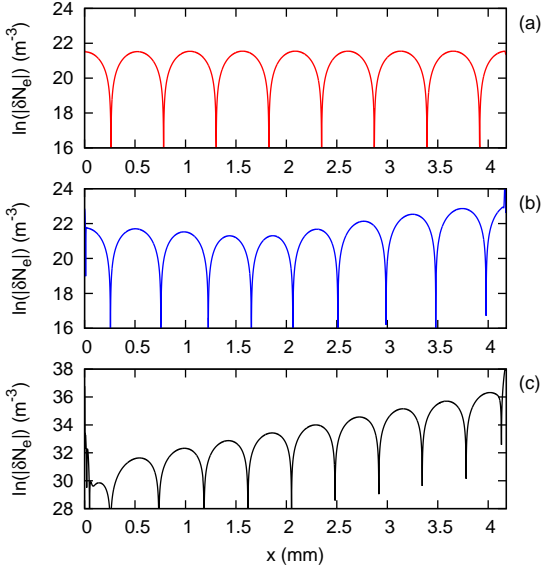}
\caption{\label{fig:02} %
Spatial profiles of bulk electron density perturbation obtained at
$t_A=0.35\text{~ns}$ (a), $t_B=3.01\text{~ns}$ (b), and $t_C=141.8\text{~ns}$
(c). Times $t_{A,B,C}$ are shown by arrows A, B, and C in Fig.~\ref{fig:01}. }
\end{figure}

The oscillations have the wavelength of the initial perturbation during only
the first few periods. The initial oscillation pattern corresponds to a
standing wave. As the instability develops, the standing wave transforms to
a propagating wave, see Fig.~\ref{fig:01}a. This process is accompanied by
the shrinking of the wavelength, compare the density perturbation profiles
at three consecutive times in Fig.~\ref{fig:02}. At the initial phase of the
instability, the perturbations propagate with the original beam velocity,
see Fig.~\ref{fig:01}a. At the asymptotic stage given by Eq.(\ref{eq:04}) with
the spatial growth rate along the beam propagation, the wave phase velocity
is noticeably lower than the velocity of beam propagation, compare the slope
of the black dashed line with that of the black solid lines in
Fig.~\ref{fig:01}b.
%
%------------------------------------------------------------------------ Fig_01.eps
%
%
%------------------------------------------------------------------------ Fig_02.eps
%

\begin{figure}
\includegraphics{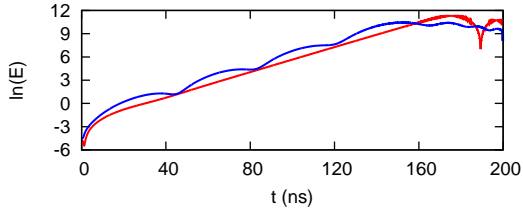}
\caption{\label{fig:03} %
Amplitude of electric field oscillations versus time in fluid simulations with
$\alpha=0.0006$ and $L=4\lambda_b$ (red curve), $L=4.7\lambda_b$ (blue
curve). The curves are obtained in the point with coordinate $x=3.55\text{~mm}$
(red) and $x=4.34\text{~mm}$ (blue). }
\end{figure}

Simulation reveals that before the asymptotic state establishes, the
temporal growth rate changes with time, see Fig.~\ref{fig:03}. Initially,
the growth rate is large compared to the analytical value defined by
Eqs.~(\ref{eq:14}) and (\ref{eq:13}). Then it gradually decreases towards the
asymptotic value predicted by the theory and it stays approximately constant
for tens and even hundreds of plasma periods until the nonlinear stage of
instability and its saturation occurs, see the red curve in Fig.~\ref{fig:03}.
Note that the modification of the wavelength mentioned above stops when
the instability reaches the asymptotic stage, which for the red curve in
Fig.~\ref{fig:03} occurs near 20 ns.
%
%----------------------------------------------------------------------- Fig_03.eps
%

In order to investigate the dependence of the growth rate on plasma
parameters, four simulation sets are discussed below. In set one, the ratio
of the beam to plasma density is $\alpha\equiv n_{b,0}/n_{p,0}=0.00015$, the
size of the system $L$ increases from $\lambda_{b}$ to $8.5\lambda_{b}$. Set
two is similar to set one but the beam density is higher, $\alpha=0.0006$. In
set three, $L=3.4\lambda_{b}$ is constant while
$\alpha$ changes from 0.0001 to 0.0006. The fourth set is similar to set
three but $L=8.3\lambda_{b}$.

\begin{figure}
\includegraphics{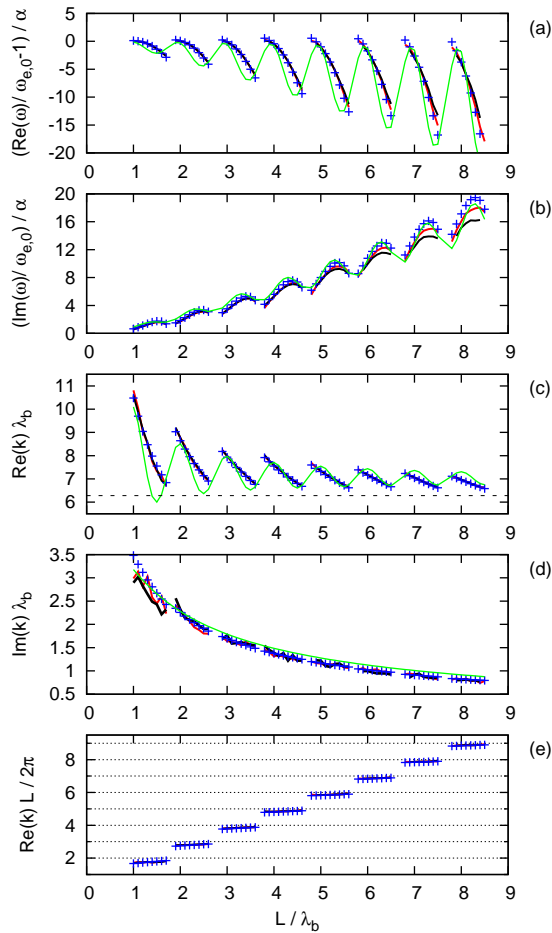}
\caption{\label{fig:04} %
Frequency (a), temporal growth rate (b), wavenumber (c),
spatial growth rate (d), and the number of wave periods per system length (e)
versus the length of the system. The blue crosses mark values obtained by
analytical solution given by equations (\ref{eq:13}), (\ref{eq:14}), and
(\ref{eq:12}). Solid red and black curves represent values obtained in fluid
simulations with $\alpha=0.00015$ (red) and $\alpha=0.0006$ (black). Solid green
curves are values provided by fitting formulas (\ref{eq:17}), (\ref{eq:18}),
(\ref{eq:19}), and (\ref{eq:20}). In (c), the black dashed line marks the
resonant wavenumber. }
\end{figure}

In all simulations, the growth rates, the frequencies, and the wavenumbers
are calculated during the asymptotic stage when the temporal growth rate is
constant for a prolonged period of time, see the red curve in Fig.~\ref%
{fig:03} for $20\text{~s}<t<160\text{~s}$. In simulation sets one and two,
for some values of $L$ such a stage never appears, see the blue curve in
Fig.~\ref{fig:03}. These values of $L$ correspond to the gaps in the
simulation data seen in Fig.~\ref{fig:04}.

Overall, there is an excellent agreement between the simulations and the
theory. The dimensionless values of $[\text{Re}(\omega)-\omega
_{e,0}]/(\omega_{e,0}\alpha)$, $\text{Im}(\omega)/(\omega _{e,0}\alpha)$,
$\text{Re}(k\lambda_{b})$, and $\text{Im}(k\lambda_{b})$ obtained in simulation
sets one and two (red and black curves in Fig.~\ref{fig:04}) and by analytical
solution of the theoretical dispersion relation (blue crosses in
Fig.~\ref{fig:04}) are very close to each other and appear to be functions of
the dimensionless system length only, as predicted by the analytical solution
given by Eqs.~(\ref{eq:12}) and (\ref{eq:14}). These functions for
$\text{Re}(\omega)$, $\text{Im}(\omega)$, and $\text{Re}(k)$ have band
structure. Mathematically, it is the consequence of the presence of many
branches in the Lambert function. The instability growth is given by the
maximum growth rate value that changes from branch to branch when the gap size
crosses some critical value, typically when $L/\lambda_{b}$ approaches an
integer, see Fig.~\ref{fig:04}. Similar band structure was also observed for
the Pierce diode.~\cite{Piel:2010, Cary:1982} Figure~\ref{fig:04}e shows the
number of wave periods in the gap as a function of the gap length. In all
cases, it is very close to an integer number, although not exactly:
%
%========================================================================= eq:16
\begin{equation}\label{eq:16}
\text{Re}(k)L/(2\pi)\simeq \lceil L/\lambda _{b}\rceil ,
\end{equation}%
where $\lceil x\rceil \equiv ceiling(x)$ is the
smallest integer not less than $x$.
%That is the wave structure is close to
%growing standing wave.
%
%Note that in fluid simulations, a small sheath forms
%near the electrodes and more accurate boundary conditions need to account
%sheath effect, as it is done for example in Ref.~\cite{Smolyakov:2013}.

Since the shape of the functions is universal for various beam densities, it
is reasonable to introduce approximate formulas which fit the numerical
solution as follows:
%
%========================================================================= eq:17
%
\begin{equation}\label{eq:17}
\text{Re}(\omega) \approx \frac{\omega _{e,0}\alpha }{18}L_{n}\ln (L_{n})\left[
1-0.9\cos \left( L_{n}+0.4\right) \right],
\end{equation}
%
%========================================================================= eq:18
%
\begin{equation}\label{eq:18}
\text{Im}(\omega) \approx \frac{\omega _{e,0}\alpha }{13}L_{n}\ln (L_{n})\left[
1-0.18\cos \left( L_{n}+\frac{\pi }{2}\right) \right],
\end{equation}%
%
%========================================================================= eq:19
%
\begin{equation}\label{eq:19}
\text{Re}(k)\approx \frac{\omega_{e,0}}{v_{b,0}} \left[ 1.1+\frac{1+2.5\cos
(L_{n})}{1.1L_{n}}\right],
\end{equation}%
%
%========================================================================= eq:20
%
\begin{equation}\label{eq:20}
\text{Im}(k)\approx \frac{\omega_{e,0}}{v_{b,0}} \frac{2\ln (L_{n})-0.5}{L_{n}}.
\end{equation}

\begin{figure}
\includegraphics{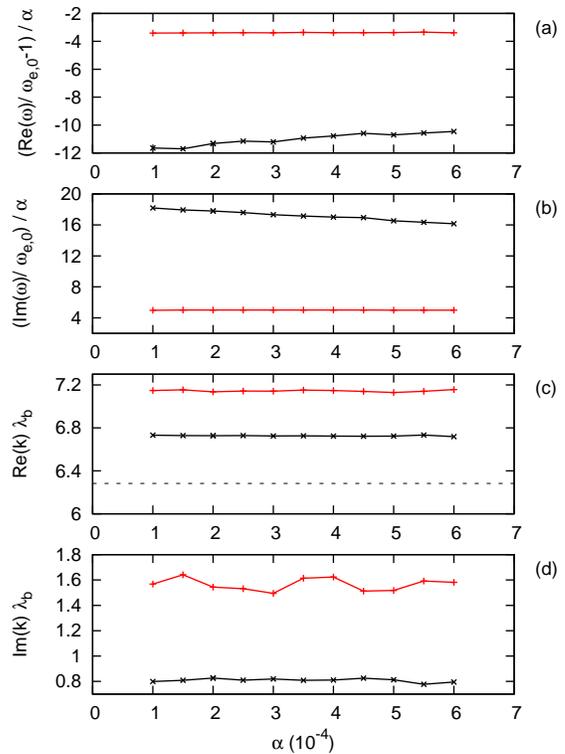}
\caption{\label{fig:05} %
Frequency (a), temporal growth rate (b), wavenumber (c), and
spatial growth rate (d) versus the ratio of the beam and bulk electron
densities in fluid simulations with $L=3.4\protect\lambda _{b}$ (red) and
$L=8.3\protect\lambda _{b}$ (black). In (c), the black dashed line marks the
resonant wavenumber. }
\end{figure}

The wavenumber and the spatial growth rate depend on the system length but
are virtually insensitive to the beam density, see Fig.~\ref{fig:05}c and
Fig.~\ref{fig:05}d. The temporal growth rate is approximately linearly
proportional to the relative beam density $\alpha $. The linear law holds
especially well for short systems, see the red curve in Fig.~\ref{fig:05}b
and compare red and black curves for $L/\lambda_{b}<6$ in Fig.~\ref{fig:04}b.
For longer systems, however, deviation from the linear law becomes
noticeable.
%
%-------------------------------------------------------------------------- Fig_04.eps
%
%-------------------------------------------------------------------------- Fig_05.eps
%

\section{Temporal growth rate in kinetic simulation \label{sec:kinetic}}

Kinetic simulations are carried out with the EDIPIC 1D3V particle-in-cell (PIC)
code.\cite{Sydorenko 2006} The code is modified to reproduce conditions of the
fluid simulations. The ions form an immobile background, the boundaries have
zero potential. The bulk electrons are reflected specularly from the
boundaries. The beam electrons penetrate through the boundaries freely. The
initial plasma density and the beam energy are the same as in the fluid
simulations. Collisions are omitted. Two simulations are carried out with
$L=8.3\lambda_{b}$, $\alpha=0.0006$ but different number of particles per cell.
One simulation has 10000 particles per cell. The other simulation has 2000
particles per cell. Below these simulations are referred to as 10k and 2k
simulations, respectively.

\begin{figure}
\includegraphics{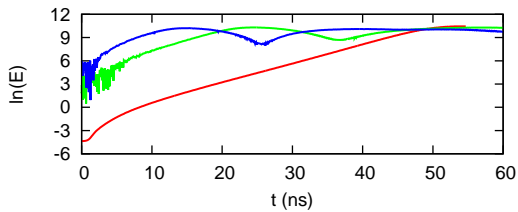}
\caption{\label{fig:06} %
Amplitude of electric field oscillations vs time in simulations
with $L=8.3\lambda_b$ and $\alpha=0.0006$. The curves
represent fluid simulation (red), PIC simulation with 10000 particles per
cell (green), and PIC simulation with 2000 particles per cell (blue). }
\end{figure}

PIC simulations start with a significant level of statistical noise which is
few orders of magnitude higher than the initial perturbation induced in the
fluid simulations above. At the same time, the amplitudes of nonlinear
saturation of the instability in PIC and fluid simulations are close to each
other. As result, the time when the oscillations grow from the initial noise
level to the saturation in PIC simulation is much shorter than that in a
fluid simulation. Moreover, at the initial stage the growth rate gradually
decreases which furthermore limits the duration of the asymptotic stage
described by analytic solution. For example, in the 10k simulation, the
asymptotic stage lasts from 10 ns to 20 ns while in the fluid simulation
that stage occurs between 10 ns and 45 ns, compare the green and the red
curves in Fig.~\ref{fig:06}. The short asymptotic stage in the 10k
simulation still allows to calculate the temporal growth rate which appeared
to be very close to the  value obtained in fluid simulations. In the 2k
simulation, however, the noise level is higher and the asymptotic stage is very
short and barely detectable, see the blue curve in Fig.~\ref{fig:06}.
%
%-------------------------------------------------------------------------- Fig_06.eps
%

\section{Saturation amplitude in kinetic simulation \label{sec:kinampl}}

%
%-------------------------------------------------------------------------- Fig_07.eps
%
\begin{figure*}
\centering
\includegraphics[scale=0.8]{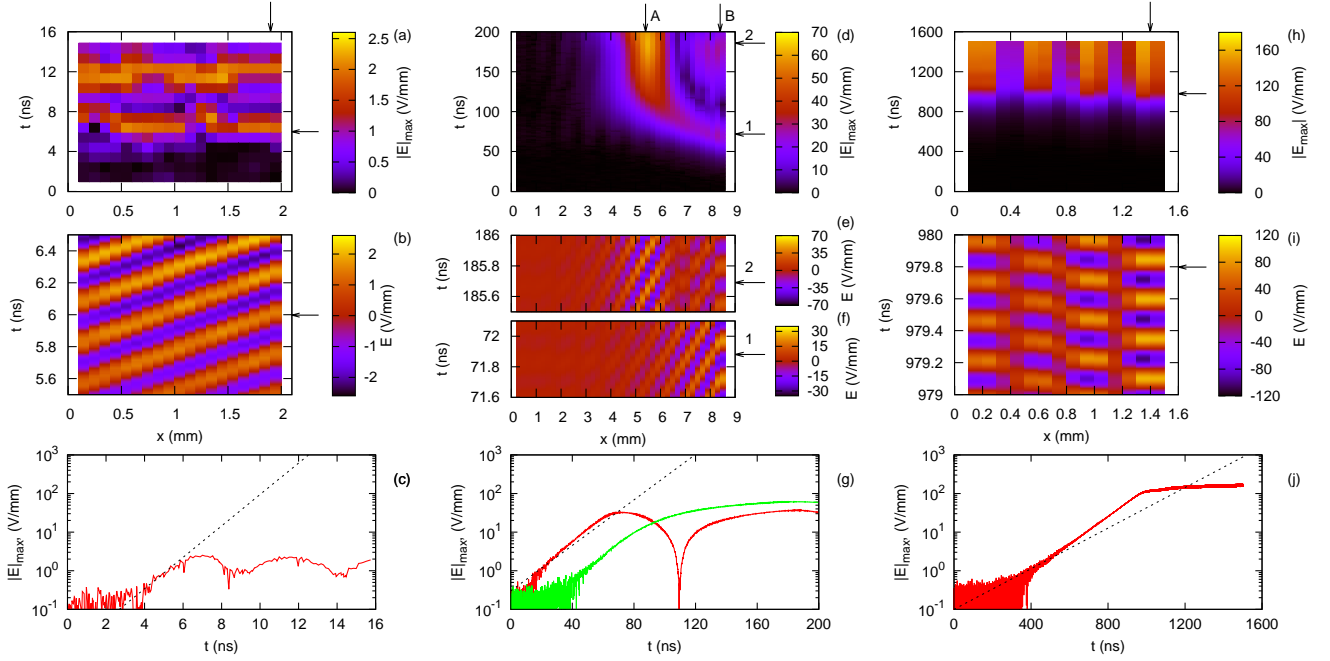}
\caption{\label{fig:07} %
Results of PIC simulation with periodic boundaries (a,b,c), non-periodic
boundaries and $L=8.3\lambda_b$ (d,e,f,g), and non-periodic boundaries and
$L=1.5\lambda_b$ (h,i,j).
The top row (a,d,h) shows amplitude of oscillations versus coordinate and time.
The middle row (b,e,f,i) shows electric field versus coordinate and time. The
bottom row (c,g,j) shows the amplitude of oscillations versus time at certain
locations marked by vertical arrows in (a), (d), and (h), respectively. The
green and the red curves in (g) correspond to locations marked by vertical
arrows A and B in (d), respectively.
The horizontal arrows in (a,b) mark time of the snapshot shown in
Figs.~\ref{fig:08}(a,b). The horizontal arrows in (d,e,f) mark
time of snapshots shown in Figs.~\ref{fig:08}(c,d) (arrow 1) and
in Figs.~\ref{fig:08}(e,f) (arrow 2). The horizontal arrows in
(h,i) mark time of the snapshot shown in Figs.~\ref{fig:08}(g,h).
The dashed black straight line in (c) shows the exponential growth with the
theoretical growth rate in an infinite plasma. The dashed black straight lines
in (g) and (j) corresponds to the growth rates obtained in fluid simulations
with the same system length and beam current. }
\end{figure*}
PIC simulations described below are carried out with the following common
parameters. The initial uniform plasma electron density is $n_{e,0}=2\times
10^{17}\text{~m}^{-3}$, the initial electron beam energy or the energy of
injection is $W_b=50\text{~eV}$, the beam-to-plasma density ratio is
$\alpha=1.5\times 10^{-4}$, the initial plasma electron temperature
$T_{e,0}=0.5\text{~eV}$, the size of a cell of the computational grid is
$\Delta x=2.078\times 10^{-6}\text{~m}$ corresponding to $\lambda_{D,e}/8$
where $\lambda_{D,e}$ is the electron Debye length of the ambient plasma, both
the plasma and the beam initially are represented by 2500 macroparticles per
each cell of the grid. The ions are represented by an immobile uniform
background with density which ensures that the plasma-beam system is initially
neutral. The electron beam propagates in the positive $x$-direction.

The following three PIC simulations are carried out. First simulation has
periodic boundary conditions and the system length of $L=2\lambda_b$ where
$\lambda_b=2\pi V_b/\omega_{e,0}$ is the wavelength of the plasma wave resonant
with the beam in an infinite plasma, $V_b$ is the beam velocity. For the
selected parameters, $\lambda_b=1.044\text{~mm}$. Second simulation has
non-periodic boundary conditions similar to the ones used in the fluid model.
The boundaries are grounded, the plasma electrons are reflected from the
boundaries while the beam electrons penetrate through them freely. System
length in the second simulation is $L=8.3\lambda_b$ which corresponds to the
maximum of the temporal growth rate in the dispersion band with
$L/\lambda_b\approx 8$ in the fluid simulation. Third simulation also has the
non-periodic similar to the second simulation, but the system length is much
shorter, only $L=1.5\lambda_b$. In the second and the third simulations, the
beam injection occurs at the boundary $x=0$.
%
%-------------------------------------------------------------------------- Fig_08.eps
%
\begin{figure*}
\centering
\includegraphics[scale=0.85]{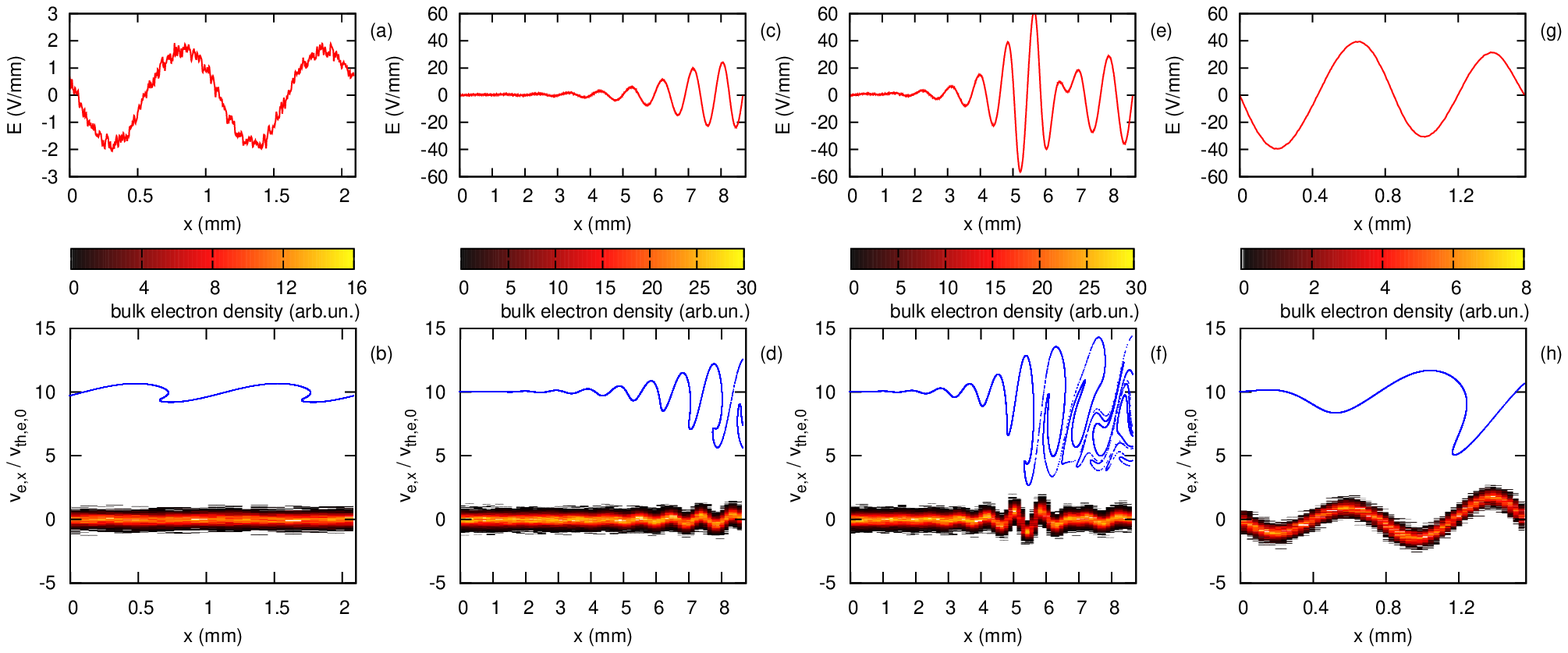}
\caption{\label{fig:08} %
Electric field profiles (a,c,e,f) and electron ``velocity versus coordinate''
phase planes (b,d,f,h) in the PIC simulations with periodic boundaries (a,b),
non-periodic boundaries and $L=8.3\lambda_b$ (c,d,e,f), and non-periodic
boundaries and $L=1.5\lambda_b$ (g,h).
Snapshots (a,b) are obtained at time 6 ns shown by the horizontal arrows in
Figs.~\ref{fig:07}(a,b). Snapshots (c,d) are at time
71.88 ns shown by arrows 1 in Figs.~\ref{fig:07}(d,f).
Snapshots (e,f) are at time 185.69 ns shown by arrows 2 in
Figs.~\ref{fig:07}(d,e). Snapshots (g,h) are at time
979.8 ns shown by the horizontal arrows in
Figs.~\ref{fig:07}(h,i).
In the phase planes, the beam electrons are represented by the blue color while
the plasma electrons are represented by the color map, the white background
correspond to the empty space. }
\end{figure*}

The theory of interaction of a cold beam with a cold plasma predicts that the
exponential growth of the amplitude of plasma oscillations is followed by
saturation and subsequent amplitude oscillations.\cite{Matsiborko:1972} Such a
picture is reproduced in the first simulation, see Figs.~\ref{fig:07}(a) and
(c). The plasma wave propagates in the direction of beam propagation and has
constant amplitude along the system, see Fig.~\ref{fig:07}(b) and
Fig.~\ref{fig:08}(a). The theoretical growth rate is
%----------------------------------------------------------------------------------- eq:21
\begin{equation}\label{eq:21}
    Im(\omega)=0.7\alpha^{1/3}\omega_{e,0},
\end{equation}
and the electric field amplitude in the first maximum is
%----------------------------------------------------------------------------------- eq:22
\begin{equation}\label{eq:22}
    E_{1,max}=3 k W_{b} \alpha^{2/3},
\end{equation}
where $\omega_{e,0}$ is the Langmuir frequency of plasma electrons and
$k=\omega_{e,0}/V_b$ is the resonance wavenumber. For the selected beam and
plasma parameters, $\omega_{e,0}=2.52\times 10^{10}\text{~s}^{-1}$,
$V_b=4.19\times 10^6\text{~m/s}$, and $k=6015.9\text{~m}^{-1}$. Therefore, the
theoretical growth rate (\ref{eq:21}) is
%----------------------------------------------------------------------------------- non-numbered
\begin{equation*}
  Im(\omega)=0.938\times 10^9\text{~s}^{-1}
\end{equation*}
and the electric field amplitude maximum (\ref{eq:22}) is
%----------------------------------------------------------------------------------- non-numbered
\begin{equation*}
  E_{1,max}= 2549 \text{V/m}~.
\end{equation*}
Both values are very close to the simulation results, see the red curve in
Fig.~\ref{fig:07}(c) and compare it with the dashed straight line which
corresponds to the theoretical growth rate (\ref{eq:21}). It is necessary to
mention here that during the first 4 ns, the growing oscillations are obscured
by the noise present in the system due to the finite number of particles in
simulation. The saturation begins at t=6 ns when the beam particles start
passing each other, see the phase plane in Fig.~\ref{fig:08}(b). Note that by
this time the beam electrons in the laboratory frame travel about 25 mm which
is several times more than the length of the system in the second simulation.

In the second simulation, the boundary conditions are non-periodic and the
linear stage of the instability follows the fluid theory for the finite-length
systems developed above -- the wave amplitude grows both along the system and
in time, the temporal growth rate is close to the fluid value, compare the red
curve with the straight black dashed line in Fig.~\ref{fig:07}(g). The
saturation of the amplitude occurs around 70 ns and here the amplitude is
maximal near the exit end of the system, see Fig.~\ref{fig:07}(d) and (f) near
arrow 1 and the electric field profile in Fig.~\ref{fig:08}(c).
The maximum wave amplitude is an order of magnitude higher than that in the
periodic simulation, and it causes much stronger velocity perturbations of the
beam particles. Due to both limited distance of interaction between the beam
and the wave and the nonuniform wave amplitude, the beam electrons start
passing each other only near the exit end, see Fig.~\ref{fig:08}(d).

An interesting process occurs after the first saturation. Position of the
maximum amplitude gradually moves towards the injection boundary until it
reaches the distance of about 5.5 mm, and the value of the maximum almost
doubles, see Fig.~\ref{fig:07}(d) and Fig.~\ref{fig:08}(e). After 150 ns, it is
this new maximum where the passing of beam electrons is achieved, not the exit
end of the system, see Fig.~\ref{fig:08}(f). Downstream of this maximum
($x>6\text{~mm}$) the beam electrons are completely mixed in the phase plane.
As a result of this change in the beam structure, while upstream of the maximum
($x<5\text{~mm}$) the wave propagates along the beam direction with spatially
growing amplitude, downstream of the maximum the wave pattern is closer to that
of a standing wave, compare Figs.~\ref{fig:07}(e) and (f).

The second simulation clearly shows that the amplitude of saturation of the two
stream instability in finite length plasmas can be significantly higher than
that in an infinite plasma. In a bounded system, the length of interaction
between a beam electron and the plasma wave cannot exceed the distance between
the boundaries. An infinite plasma has no such a limit. Here a wave of modest
intensity can interact with beam electrons over longer distances before the
mixing of the beam electrons in the phase plane occurs. In order to achieve
such a mixing over much shorter distances, which is the case in bounded
systems, the wave field must be much stronger.

In order to find how strong this effect can be, the third simulation with the
non-periodic boundary conditions is performed with a very short system length
$L=1.5\lambda_b$. The wave pattern corresponds to a standing wave with 5 nodes
and 4 antinodes. The antinodes and the 3 middle nodes are clearly visible in
Figs.~\ref{fig:07}(h) and (i). There are two nodes with approximately zero
electric field at the ends of the system. These two nodes were not resolved by
the code diagnostics used to produce Figs.~\ref{fig:07}(h) and (i), but they
are visible in the electric field profile in Fig.~\ref{fig:08}(g). The temporal
growth rate in this simulation is about 34\% higher than the growth rate in the
fluid simulation with the same parameters, compare the red curve with the black
straight dashed line in Fig.~\ref{fig:07}(j). The maximal wave amplitude
reaches 170 V/mm which is \textit{more than 60 times stronger than the field in
the periodic system}, compare Figs.~\ref{fig:07}(j) and (c). Such a strong
electric field produces mixing of beam electrons on a very short distance of 1
mm which is about one resonance wavelength, see Fig.~\ref{fig:08}(h).
%
%-------------------------------------------------------------------------- Fig_09.eps
%
\begin{figure}
%\centering
\includegraphics[scale=1.0]{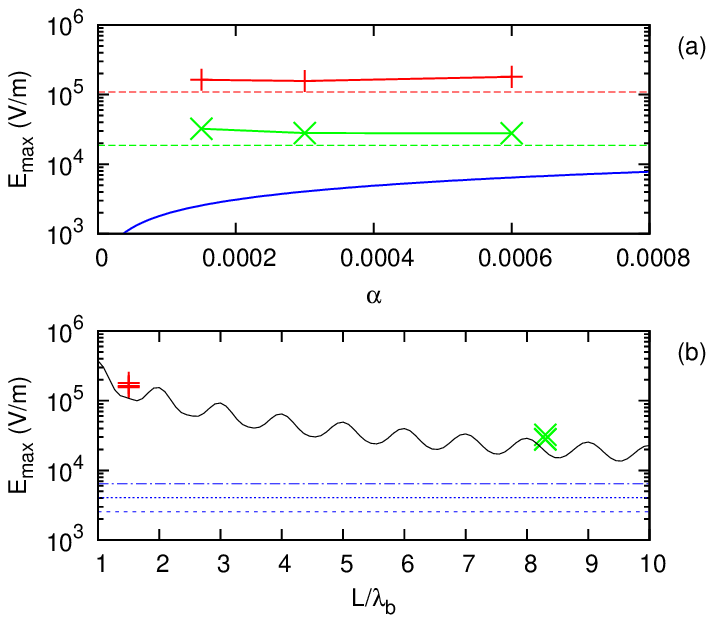}
\caption{\label{fig:09} %
(a) Amplitude of the first maximum versus the relative beam density in PIC
simulations with $L=8.3\lambda_b$ (green) and $L=1.5\lambda_b$ (red). In (a),
the solid blue curve is the theoretical prediction (\ref{eq:22}) for an
infinite plasma, the dashed horizontal lines mark theoretical predictions given
by Eq.~(\ref{eq:25}) for finite-length systems with $L=8.3\lambda_b$ (green) and
$L=1.5\lambda_b$ (red).
(b) Saturation amplitude versus the system length. In (b), the black curve is
obtained with Eq.~(\ref{eq:25}), the red vertical crosses and the green diagonal
crosses mark first amplitude maxima in the PIC simulations with
$L=8.3\lambda_b$ (green) and $L=1.5\lambda_b$ (red); the blue horizontal lines
mark theoretical values (\ref{eq:22}) for the infinite plasma with the beam of
relative density $\alpha=0.00015$ (dash), $0.0003$ (short dash), and $0.0006$
(dash-dot).
The PIC simulation values for $L=8.3\lambda_b$ and $L=1.5\lambda_b$ are
obtained at points marked by vertical arrow B in
Fig.~\ref{fig:07}(d) and by the vertical arrow in
Fig.~\ref{fig:07}(h), respectively.
}
\end{figure}

The phase plots shown in Fig.~\ref{fig:08} prove that in the finite length
system the saturation of the instability occurs when the beam particles are
overtaking each other. This process depends on the wave amplitude and the
system length but should not depend on the beam current. To check this, two
additional simulations are carried out with $L=8.3\lambda_b$ and the relative
beam density of $\alpha=0.0003$ and $0.0006$. Another two additional
simulations with these beam densities are carried out for $L=1.5\lambda_b$. The
results of these simulations combined with those obtained above for
$\alpha=0.00015$  confirm that the amplitude of the first maximum of saturation
of the instability is virtually insensitive to the beam current, see the red
and the green curves with markers in Fig.~\ref{fig:09}(a). The only difference
is that for higher current the saturation is achieved faster. Note that in the
whole range of the beam density $\alpha$ considered, the values of the
saturation amplitude in the finite system are much higher than the predictions
for the infinite system, compare the red and the green curves with markers with
the blue curve in Fig.~\ref{fig:09}(a).

The wave amplitude which causes overtaking of beam particles can be estimated
as
%----------------------------------------------------------------------------------- eq:23
\begin{equation}\label{eq:23}
    \dfrac{eE}{m_e(\omega-kv_b)^2}\sim\lambda_b
\end{equation}
where the left-hand side is the displacement of the particles trapped by the
wave in the wave frame. Using equation (\ref{eq:12}) one can replace
$(\omega-kv_b)^2$ with $(\chi\omega_{e,0})^2$. Then, replacing
$\lambda_b\omega_{e,0}$ with $v_{b,0}$ one can write an expression for the
maximal electric field of the wave as
%----------------------------------------------------------------------------------- eq:24
\begin{equation}\label{eq:24}
    E_{max}=\dfrac{m_e v_{b,0} \omega_{e,0}}{e}|\chi^2|~.
\end{equation}
For estimates, the value of $\chi^2$ is convenient to find as
%----------------------------------------------------------------------------------- eq:25
\begin{equation}\label{eq:25}
    \chi^2=\left\{1-\dfrac{v_{b,0}}{\omega_{e,0}}[Re(k)+i Im(k)]\right\}^2
\end{equation}
with $Re(k)$ and $Im(k)$ given by the approximate formulas (\ref{eq:19}) and
(\ref{eq:20}). Note that expressions (\ref{eq:19}) and (\ref{eq:20}) are
independent on beam current and are functions of the normalized plasma gap
width $L_n$ only. Therefore, the maximal field (\ref{eq:24}) depends on the
beam velocity $v_b$ and the gap width $L$ but does not depend on the beam
current.

A dependence $E_{max}(L)$ calculated with (\ref{eq:25}) for the beam parameters
used in the simulations above is shown by the black curve in
Fig.~\ref{fig:09}(b). The oscillations in this curve reflect the band structure
of the wave number in the finite length system. The saturation values obtained
in the PIC simulations are remarkably close to the values given by
Eq.~(\ref{eq:25}), compare curves with markers with the horizontal dashed
curves of the same color in Fig.~\ref{fig:09}(a), also compare the markers with
the black curve in Fig.~\ref{fig:09}(b). The value of $E_{max}$ decreases with
$L$ and eventually approaches the saturation values for the infinite system
given by Eq.~(\ref{eq:22}), compare the black curve with the horizontal blue
lines in Fig.~\ref{fig:09}(b).

PIC simulations discussed above demonstrate that the growth of the maximal
electric field in the two-stream instability in a short system compared to an
infinite plasma can be very large. It is necessary to mention, however, that
these simulations are carried out with certain simplifications similar to those
made in the fluid model. In particular, the sheath is not resolved, the ion
background is immobile, and collisions with neutrals are omitted. The realistic
sheath will allow some energetic plasma electrons to escape and may affect the
structure of the wave interacting with the beam. If the ion dynamics is
accounted for, the strong plasma oscillations may result in the modulation
instability which will create density cavities and affect the wave. Finally,
the two stream instability can be suppressed by electron-neutral collisions if
the collision frequency is more than two times the collisionless growth rate.
In very short systems, the temporal growth rate is very small, which means that
the neutrals present in a real beam-plasma system may simply prevent the
instability from developing.

\section{Summary \label{sec:summary}}

In summary, we have studied the development of the two-stream instability in a
finite size plasma bounded by electrodes both analytically and making use of
fluid and particle-in-cell simulations. We show that the instability reaches
the asymptotic state when the wave structure has the same spatial profile and
grows in time with a constant growth rate. The spatial structure of the wave is
close to a standing wave but has a spatial growth along the beam propagation.
We derived analytic expressions (\ref{eq:17}-\ref{eq:20}) for the frequency,
wave number and the spatial and temporal growth rates. Obtained analytic
solution agrees well with the values given by fluid and particle-in-cell
simulations.

The saturation of the instability occurs due to the overtaking of beam
particles. Formulas for the estimate of the saturation amplitude
(\ref{eq:24}-\ref{eq:25}) are derived and are in good agreement with the
simulation results. The amplitude of saturation does not depend on the beam
current but grows significantly for shorter systems. \textit{Compared to the
value predicted for an infinite plasma, the saturation amplitude for
low-current plasma beam systems of length of a few resonance wavelengths may be
higher by more than an order of magnitude.}

\textit{Acknowledgement:} This research was supported in part by U.S.
Department of Energy and Air Force Office of Scientific Research. Authors
acknowledge valuable discussions with Edward Startsev and Peter Ventzek.

\end{document}